\documentclass{PoS}
\usepackage{lineno}
\usepackage{xcolor}
\usepackage{amsmath}
\title{Effects of Lorentz invariance violation on cosmic ray photon emission and gamma ray decay processes}

\ShortTitle{Effects of LIV on cosmic and gamma rays}

\author{
\speaker{H. Mart\'inez-Huerta}$^{\rm ~a,b}$ 
and 
A. P\'erez-Lorenzana$^{\rm b}$ 
\thanks{HMH thanks the support via FAPESP (2017/03680-3). This work was partially supported by Conacyt grant No.~237004. }\\

        $^{\rm a}$ Instituto de F\'isica de S\~ao Carlos, Universidade de S\~ao Paulo,\\
        $\ \ \ $Av. Trabalhador S\~ao-carlense 400, S\~ao Carlos, Brasil.\\
        $^{\rm b}$Departamento de F\'isica, Centro de Investigaci\'on y de Estudios Avanzados del I.P.N.,\\ 
        $\ \ \ $Apdo. Post. 14-740, 07000, Ciudad de M\'exico, M\'exico.\\
        
        $~~$E-mail: \email{humbertomh@wifsc.usp.br,  aplorenz@fis.cinvestav.mx}
        }

\abstract{
In this work, we use Lorentz invariance violation (LIV) introduced as a generic modification to particle dispersion relations to study some consequences of single photon emission, known as vacuum Cherenkov radiation, and photon decay processes in cosmic and gamma rays. These processes are forbidden in a Lorentz invariant theory but allowed under the hypothesis of LIV. We~~show that the emission rate have a dependency on the cosmic ray primary mass and the electric charge that could modify the UHECR spectrum. Furthermore, LIV dramatically enhances photon decay into an electro-positron pair above certain energy threshold. This last effect can then be used to set limits to the LIV energy scale from the direct observation of very high energy cosmic photon events by telescopes of gamma-rays.
}

\FullConference{35th International Cosmic Ray Conference $–$ ICRC217$-$\\
		10-20 July, 2017\\
		Bexco, Busan, Korea}

\begin{document}

\section{Introduction}

Lorentz  invariance  (LI)  plays  a  central  role  in  relativity  and  also  in  the  models  that describe the fundamental forces of nature, which so far has shown to have 4: electromagnetism, weak, strong  and  gravity.   The first three are very well described by the Standard Model (SM) of particles, while General Relativity describes gravity. Both, although fundamentally di\-ffe\-rent, have no contradiction in their predictions,  nevertheless, there are several models of quantum gravity that seek to unify both fundamental theories in a single one (see for instance and Refs. within ~\cite{Bluhm,Pot,ALFARO,QG1,QG2,QG3,QG4,QG5,Gian}). 
The resulting new physics may involve new properties, such as the possibility that the laws of relativity may not be exact, which may derive in a certain violation of LI that can effectively manifest itself up to some energetic scale.
 Thus, it can be expected that  the effects of Lorentz invariance violation~(LIV) increase with energy for some energetic regime.

Following this line of thought, the most energetic phenomena known so far are  Cosmic Rays (CR), where  the primary has shown to be able to reach energies of some decades of EeV  \cite{GZK}. Thus, among the many particulars that their study and understanding reveal, they provide an interes\-ting window to test fundamental physics such as the search for signatures of LIV.  
In addition to the energetic window, the very long distances that CR propagates can lead to a significant LIV effects due to accumulative processes, for instance: photon decay, photon splitting and spontaneous photon emission (or vacuum cherenkov radiation). They are kinematically forbidden processes by LI energy-momentum conservation but can be permitted in a LIV scenarios. For the present study we have chosen to explore photon emission and decay since they can lead to extreme scenarios in cosmic and gamma ray propagation.

The LIV introduced as the modifications to the  particle dispersion relation are common in astroparticle physics literature.  
This generalization is perhaps frequently used since it could offer a simple {\it ansatz} of the derived physics from the LIV  hypothesis, moreover, it is not necessarily bound to a particular LIV-model, which allows to generalize to some point the search of LIV-signatures. 
However, this generalization can be compatible with some particular models. 
Proposals as the Standard Model Extension (SME) and some effective field theories have been working to develop a proper theoretical description that includes Lorentz and/or CPT violation\footnote{The conservation of the consecutive application of Charge conjugation, Parity transformation, and Time reversal symmetry (CPT) in a quantum field theory is a consequence of Lorentz invariance, nevertheless, not every LIV involves a violation of the CPT conservation.}. The SME \cite{SME} proposes the inclusion of all the contributions to the SM that violate Lorentz symmetry by spontaneous symmetry breaking (SSB). The minimum set of such operators that maintain gauge invariance and power-counting renormalizability conform the so-called mSME.
 
In previous works \cite{HMH-APL,Proc2,Proc3} we have derived  rates for photon emission and  decay for the generic approach that reproduces some characteristics and results based on the photon sector of the SME. In this note we provide the highlights of our results for further potential phenomenological studies that could include dedicated  analysis with current data from the observatories of cosmic and gamma-rays. Thus, in the next section we introduce the LIV model. In section \ref{sec:PE}, we present the generic approximation for the spontaneous photon emission process and we compare it with the rates obtained by the isotropic modifications of the CPT-even or modified Maxwell theory (modM) \cite{TWO-SIDE,Klin_2016,VCR-17, Klin-ICRC} and  the CPT-odd modified photon theory or Maxwell-Chern-Simons (MCS) theory \cite{VCR-MCS}. In section \ref{sec:PD} we applied the previous approximation to the photon decay process to present stringent limits to the LIV energy scale and finally, in the last section, we present our conclusions. 

\section{Generic Lorentz invariance violation}\label{sec:LIV}

A phenomenological generalization of the LIV-induced modifications to the dispersion relation formalism converges to the introduction of an extra term in the dispersion relation of a single particle. We name this generalization {\it Generic LIV mechanism} since the extra term can be motivated by the introduction of a not explicitly Lorentz invariant term at the free particle Lagrangian \cite{GLASHOW}. 
Although it is not necessarily bounded to particular LIV-model it can be compatible with some particular cases. The extra term is restricted by a dimensionless coefficient, which we call $\epsilon$.
The most general modification to the dispersion relation would rather involve a general function of energy and momentum $(A)$, such that one could write\footnote{Natural units are used  $ \hbar = c = 1$ and Minkowski metric $(1,-1,-1,-1).$  }
\begin{equation}\label{eq:DR}
    E^2-p^2 + \epsilon(A)A^2 = m^2.
\end{equation}
Demanding that the LIV term can only be significant at high energies, the coefficient  $\epsilon$ should be small, which guarantees that the physics at low energies remains Lorentz invariant. 
In order to study the underlying physics, one keeps one term of order $n$ at once from the series expansion of Eq. (\ref{eq:DR}). Assuming it as the leading term, the dispersion relation  becomes
\begin{equation}\label{eq:DR2}
	E_a^2-p_a^2 \pm \alpha_{a,n} A_a^{n+2} = m_a^2,
\end{equation}
where the sub-index $a$ stand for the $a$-particle species with four-momenta $(E_a,p_a)$. The $\epsilon^{(n)}$ derivatives are assumed over a scale factor $M$, thus the term $\alpha_{a,n}~=~\epsilon_a^{(n)}/M^n = 1/(E_{LIV}^{(n)})^n$ defines the energy scale associated to the LIV physics.  $E_{LIV}$ is commonly  associated  with the energy scale of Quantum Gravity, $E_{QG}$,  which is expected to be close to the Planck scale, $E_{Pl}\sim 10^{19}GeV$.  However, without loss of generality, the LIV correction can be just named $E_{LIV}$, as we do in the section \ref{sec:PD}. In the literature, one can find upper limits to $E_{LIV}^{(n)}$ obtained by different techniques and approaches, even beyond the Planck scale, see for instance and Refs. within 
\cite{HMH-APL,GRB-LIV,FERMI-LIV,MULTI-TEV,MAGIC-ICRC,E1,E2}.

Worth to mention that he dispersion relation in Eq. (\ref{eq:DR2}) could be compatible with the Colleman and Glashow approach in Ref. \cite{GLASHOW} for the n=0 case.
Similarly, the case $n=1$ of Eq. (\ref{eq:DR2}) stands for the one derived in Ref. \cite{Example_n1}. For a n-order model, see for instance Ref. \cite{FERMI-LIV}. 
Additional examples can be found in Refs. 
\cite{DIS1,DIS2,DIS3, JACOB, GUNTER-PH,GUNTER-PD, Stecker2009, Stecker2009NJ, GLASHOW_97}.


\section{Cosmic Ray spontaneous photon emission}\label{sec:PE}

In a LI theory, Cherenkov radiation happens when a charged particle moves faster than the phase velocity of light in a medium, then, molecules in the medium are polarized through the charged particle path and they radiate coherently. Similarly, in spontaneous photon emission or vacuum Cherenkov radiation process, the LIV-vacuum acts as an optical medium with a non-trivial refractive index, hence, particles with energy above certain energy threshold are allowed to spontaneously radiate
\cite{TWO-SIDE,VCR-MCS,GLASHOW,GLASHOW_97,CROSS,   VCR_Sch}. 

In this work the spontaneous photon emission $a_p \rightarrow a_{p'}\gamma$ , may correspond in general to leptons, protons and heavy nuclei. The subscript denotes the momentum notation and hereafter, ($\omega$,~$k$) stands for the photon four momenta components, whereas prime and not prime notation  stand for the spin 1/2 charged particle involved in the process, as indicated.
In addition, since we focus our work  to aim a phenomenology approach  for the photon sector we are not to assume the simultaneous presence of LIV corrections for fermions along the following sections. Thus, applying Eq.~(\ref{eq:DR2}) to photons, we write the generalized dispersion relation: 
%
    $\omega^2~ = k^2 (1+\alpha_n k^n).$
To proceed with this approximation, we corrected the squared amplitude probability of the process, $|M|^2$, computed from the standard QED rules but making use of the correction in the photon four-momenta. We also assume Dirac gamma matrices standard properties under a LI theory. Since the correction is taken at first LIV order correction,  any term 
$\geq \alpha_{n}^2$ is negligible and the Ward identity is preserved. With such considerations,
\begin{equation}\label{eq_M}
	\frac{1}{2}\sum_{spin} |{\rm M}|^2 = e^2 |4m_a^2 - \alpha_{n} k^{n+2} |,
\end{equation}
where $e$ is the charge of the $a$-particle. The absolute value is taken in order to ensure physical congruence due the limits and the generality in the sign of $ \alpha_{n}$, for a more detailed derivation see Ref. \cite{Proc2}.
Therefore, the spontaneous photon emission rate for any order $n$, at leading order of $\alpha_n$ and for the ultra-relativistic regime ($E,p\gg m$), is given by 
{\small{
\begin{equation}\label{eq:GamaVCR}
	\Gamma_{a\rightarrow a\gamma}^{(n)} =   \dfrac{e^2}{4\pi} \dfrac{1}{4E_a} 
 	\int_0^{\theta_{max}}  \sum_{k_i}\frac{|4m_a^2 - \alpha_n k_i^{(n+2)}| }{\omega(k_i,\alpha_{n} )} 
	   \frac{k_i^2 \sin\theta d\theta}{| p_a \cos\theta -k_i  - \frac{(1+\frac{2+n}{2}\alpha_n k^n_i)}{\sqrt{1+\alpha_{n}k^n_i}} \sqrt{k_i^2 + E_a^2-2k_ip_a\cos\theta}|},
\end{equation}}}
where $\theta$ is the emission angle and $k_i$ are the non zero photon momenta modes from the corrected energy-momenta conservation, which are found by solving the following expression for a given $n$:
\begin{equation}\label{eq_VCR_g}
	\begin{aligned}
	\left[p^2 \cos^2 \theta - (p^2+m^2)\right] k - & (p^2+m^2) \alpha_n k^{n+1}  
	+ p~\cos\theta~\alpha_n k^{n+2}=0.
	\end{aligned}
\end{equation}

\begin{figure}[ht!]
	\begin{minipage}{18pc}
	    {\centering
		\includegraphics[width=1.\textwidth]{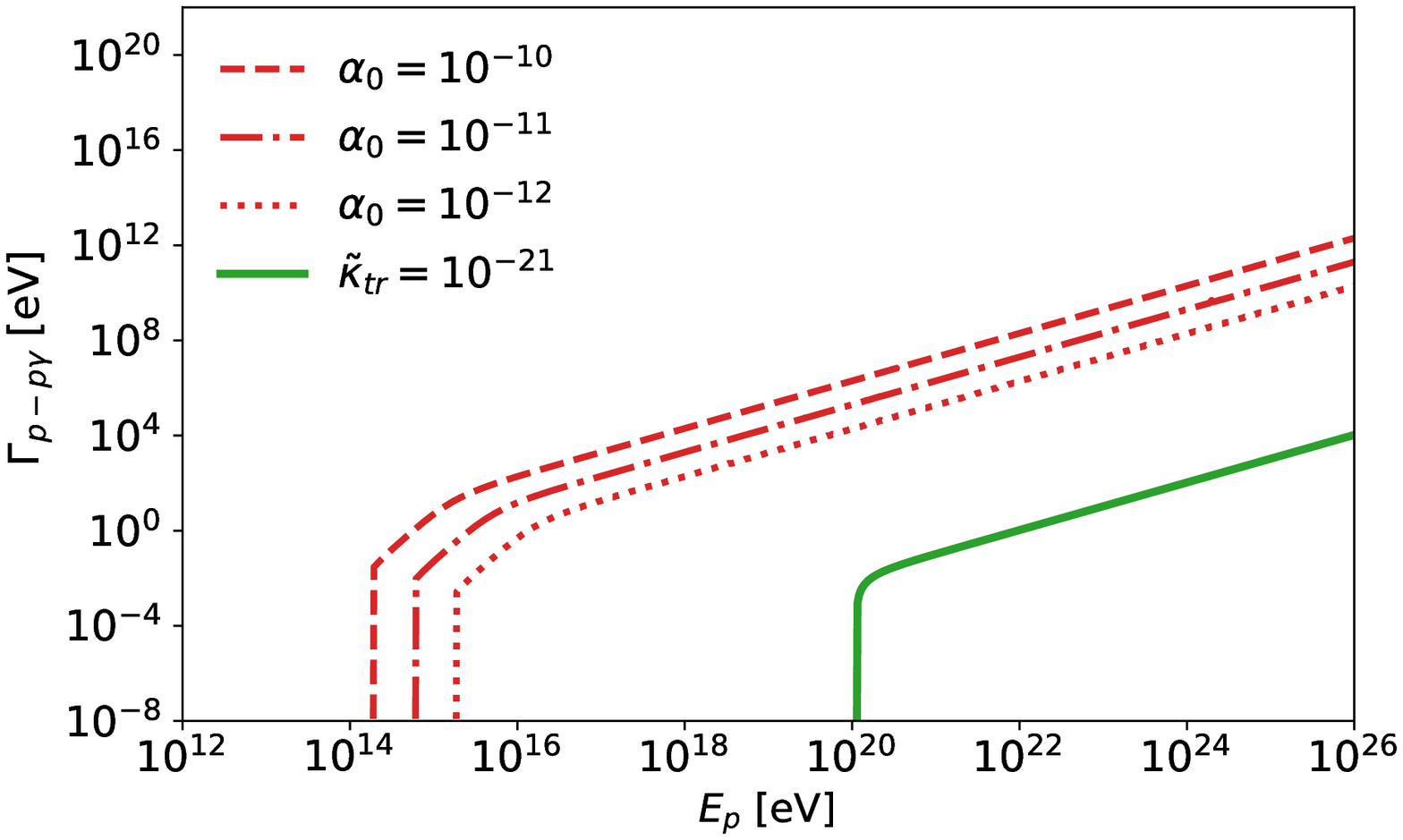}
		}
	\end{minipage}\hspace{0.5pc}%
	\begin{minipage}{18pc}
	    {\centering
		\includegraphics[width=1.\textwidth]{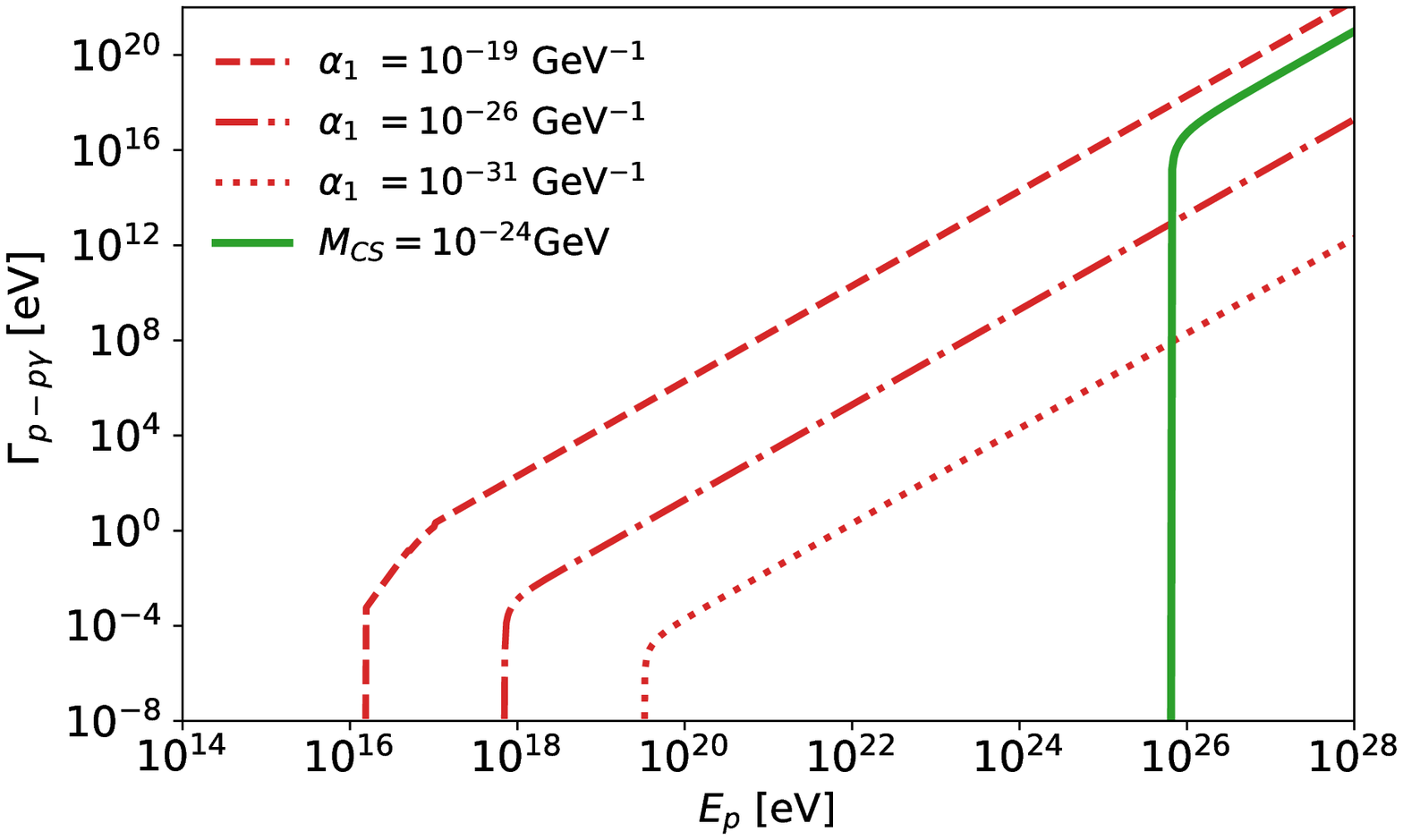}
		}
	\end{minipage} 
	\caption{Comparison of emission rates for vacuum Cherenkov radiation for a proton. Free LIV parameters were chosen as indicated. In the left we compare the case n=0 and in the right it is shown the n=1 case. All of them present a threshold which turn on the process, and a emission rate that grows with the energy but is lower for lower LIV parameter values. The slope for each n-case is compatible. }\label{fig:rates}
\end{figure}

Numerically performing the integration in Eq. (\ref{eq:GamaVCR}), we have plotted in Fig.~\ref{fig:rates}  the generic emission rate for the case $n=0$ (left) and $n=1$ (right) for different values of $\alpha_n$, for a very a small~$\theta_{max}$~($\sim10^{-2}$) and with a phenomenological threshold given by 
    $E_a \gtrsim \left(4m_a^2 / \alpha_n \right)^{1/n+2}.$
As reference and for the purpose of this work, we use $m=m_{p}$, the proton mass. For phenomenological considerations, we take the proton  nuclei as a spin 1/2 fermion.  
In the continuous (green) line, we have also plotted for comparison, the emission rate at the first order from the expansion on $E$ from  the isotropic CPT-even modified photon theory, modM (left) and the CPT-odd, MCS theory (right) from the SME. 
From a phenomenological point of view, the rate has four main characteristics. First,  it has  a drop ending energy threshold for the model,  inversely proportional to the LIV coefficient that protects the LI physics at energies below it. Second, it has a growing behaviour with the charged particle energy, noticed that, in each case, the rate grows with the same ratio. 
Third, for a given energy above the threshold, the emission rate decreases for smaller values in the LIV parameter. And fourth, it has a sensitive behaviour with the mass and charge of the emitting particle.  

\begin{figure}[ht!]
	\begin{minipage}{18pc}
	    {\centering
		\includegraphics[width=.9\textwidth]{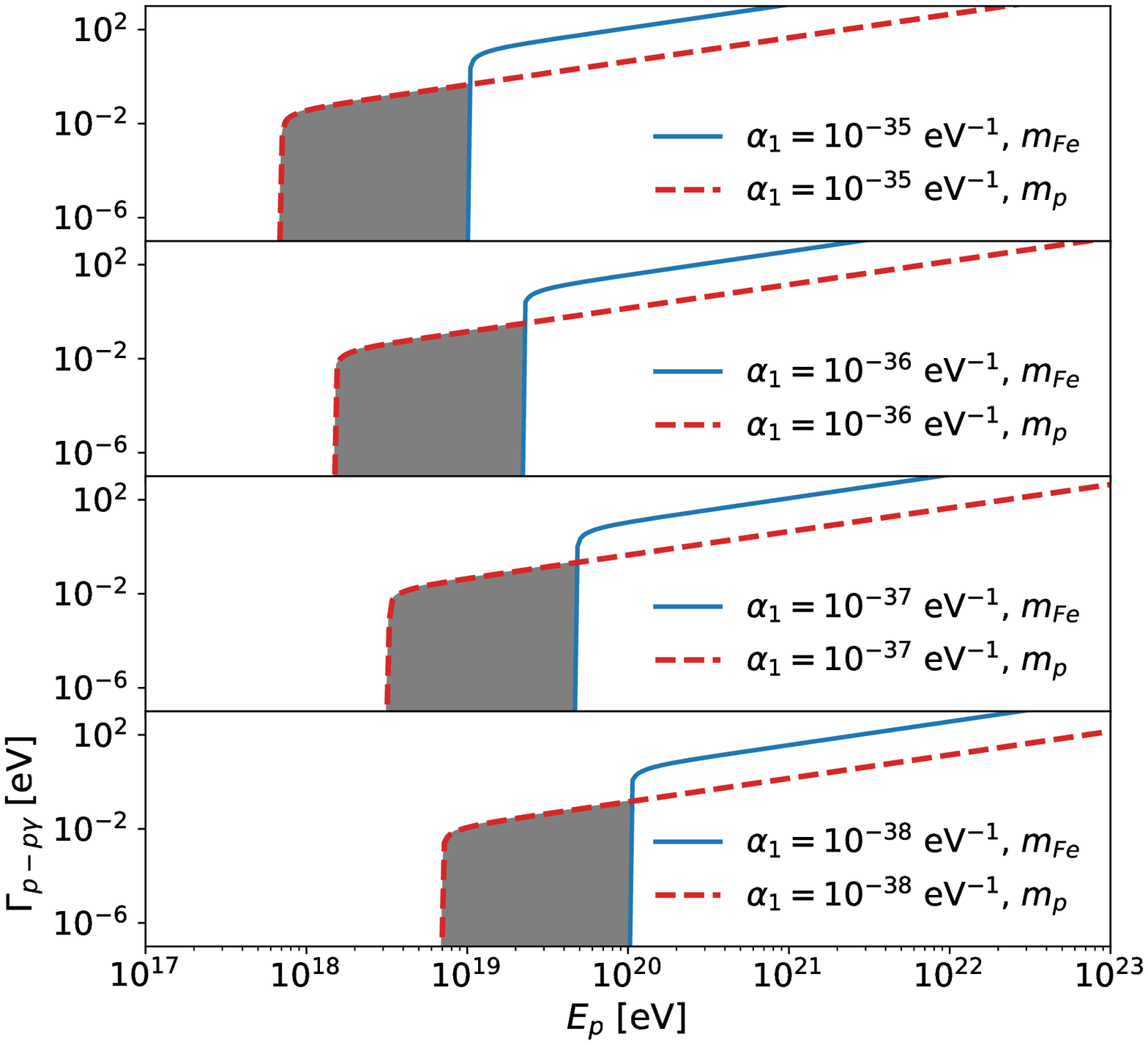}
		}
	\end{minipage}\hspace{0.5pc}%
	\begin{minipage}{18pc}
	    \centering\includegraphics[width=.85\textwidth]{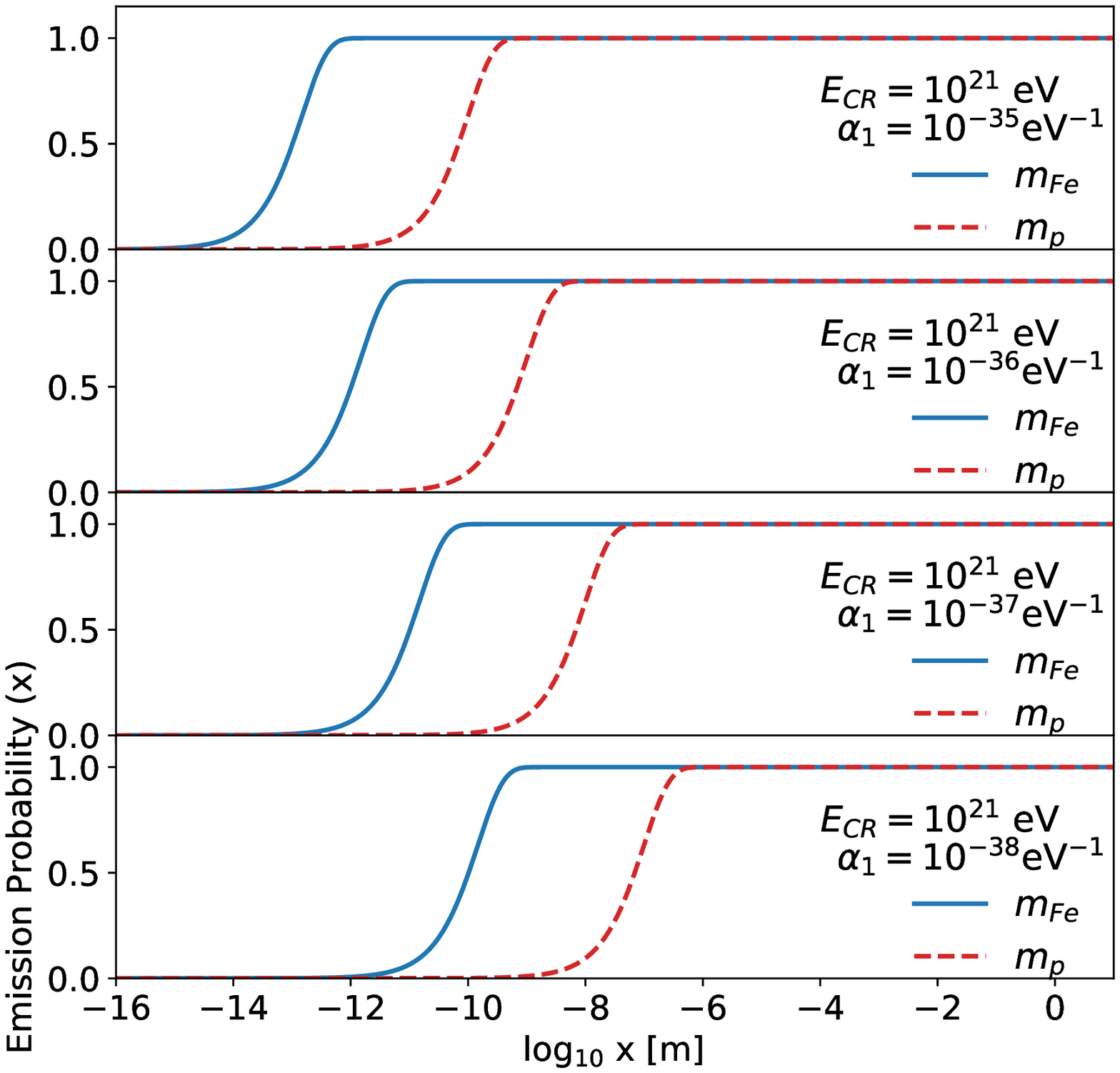}
	\end{minipage} 
	\caption{In the left, Vacuum Cherenkov energetic region, between proton and iron nuclei threshold, for different values of  n=1. A reduction to the CR flux could be expected in this region due to vacuum Cherenkov radiation that starts with the lighter components and continues with heavier.  In the shadow area the energetic region between the proton and iron nuclei threshold for different values of $\alpha_1$. In the right first emission probability as a function of mean free path.} \label{fig:win}
\end{figure}

To illustrate last feature and by considering the application of these models to the phenomenon of cosmic rays, we choose $m=m_p$ and $m=m_{Fe}$ as  representative  masses  for  a  light  and  a  heavy  component  of  the  cosmic  ray  flux, respectively. It  is  estimated  that  the  composition  of  ultra  high energy cosmic rays (UHECR) is mixed between these two limits \cite{Vitor-ICRC}. Thus, in Fig.~\ref{fig:win} (left), the  behaviour  of  the  decay rate  function  for proton and iron nuclei are compared for different values of $\alpha_1$. 
It can be seen that the effect of spontaneous photon emission is sensitive to the composition of cosmic rays. As a consequence of this attenuation process, there is a energetic region where this process only affects the lighter nuclei but not the heavier, until the energy is high enough that the process can attenuate simultaneously all the massive nuclei, that takes place in the shadow energy region.  
That is, a reduction to the CR flux could be expected in this region due to vacuum Cherenkov radiation that starts with  the  lighter  components  and  continues  with  heavier,  hence,  {\it a  tendency  to the  heavy  components  could  be  expected  in  the  UHECR  flux  located  between  both thresholds}. This effect is followed by a reduction on the contributions from the entire mass  spectrum  of cosmic rays at  higher  energies, which means that {\it forward the shadow region,  heavier  nuclei  are  attenuated  faster than lighter}, because heavier nuclei suffer the spontaneous photon emission attenuation some orders of  magnitude  higher  than  lighter  nuclei  in  a  energy  scale. To show this behaviour, in Fig. \ref{fig:win} (right) we plot first emission probability as a function of distance. 

As an example, for a value of the LIV photon parameter around $10^{-35} eV^{-1}$, one could expect that this trend would show up around an energy range of about $10^{18}$ to $10^{19}$~eV. 
UHECR observatories such as Pierre Auger, Telescope Array and HiRes have measured and reported the spectrum of ultra-energy cosmic rays at EeV energies \cite{Auger-Mix,AugerTA-Mix,HiRes-Mix}. Thus,  a dedicated analysis with their data could probe such effects or at least set limits to $\alpha_n$.

\section{Gamma ray decay process}\label{sec:PD}

Likewise the previous process, photon decay $\gamma \rightarrow a^+ a^-$, is kinematically allowed by LIV-hypothesis \cite{Klin_2016,Klin-ICRC,GLASHOW,MULTI-TEV,GUNTER-PH,GUNTER-PD,GLASHOW_97}. This decay is of special interest for gamma-ray physics since their extreme consequences makes it possible to attain tight upper limits for the generic LIV coefficient in the photon sector by using the highest photon energy astrophysical measurements.

Following the previous approximation and notation, the generic LIV photon decay rate for any order $n$, at leading order in the LIV-parameters and for a very high energy photon scenarios, follows the expression 
\begin{equation}\label{eq_PD_gamma}
	\begin{aligned}
	\Gamma_{\gamma\rightarrow a^+a^-}^{(n)} =   \dfrac{e^2}{4\pi}  \frac{|4m_a^2 - \alpha_{n}k^{n+2}| }{4 \omega(k,\alpha_{n})} 
	\int_0^{\theta_{max}} \sum_{p=p_\pm} \frac{p^2\sin\theta d\theta }{| (k\cos\theta - p)E_e - p\sqrt{k^2 + E_e^2-2kp\cos\theta}  | }, 
	\end{aligned}
\end{equation}
where the momenta modes from the corrected energy-momenta conservation and for any $n$ are
{\small
\begin{equation}\label{disc}
	p_{\pm}= \frac{1}{2(\alpha_nk^n + \sin^2\theta)}\left( \alpha_n k^{n+1}\cos\theta 
	 \pm \sqrt{\alpha_n^2 k^{2n+2}\cos^2\theta-4(\sin^2\theta+\alpha_n k^n) (1+\alpha_nk^n)m_a^2} \right) .
\end{equation}\label{fig:PDmodes}
}
Results for fixed $n$ and different $\alpha_{n}$ values can be found in \cite{Proc2,Proc3}.
The resulting rate shows to be very fast and effective once the process is allowed and suggest an abrupt cutoff in the gamma ray spectrum, thus none high-energy photons will reach the Earth from cosmological distances \cite{HMH-APL}. We have also shown that in order to ask in Eq. (\ref{fig:PDmodes})
for the momenta modes  to be real and positive, the discriminant of $p$ must be real and positive too. This condition restricts the possible emission angles for any given momentum of the photon, $\theta_{max}$, with regard to LIV parameters. Therefore, in the limit $\theta_{max}\rightarrow 0$ the process is forbidden and lower limits for $E_{LIV}$ in the photon sector can be obtained by
\begin{equation}\label{eq_limit}
    E^{(n)}_{LIV} > k_{obs}\left[\frac{k_{obs}^2-4 m^2}{4 m^2}\right]^{1/n},
\end{equation}
and the direct measurements of the highest energy photon events.
The implementation of Eq. (\ref{eq_limit}) had set stringent limits by the HEGRA and HESS measurements\cite{HEGRA, HESS}. With HEGRA event measurments we have reported $1.5\times10^{20}$~GeV and $2.8\times10^{12}$~GeV for $n=1$ and $2$ respectively \cite{HMH-APL}. Although, stringent limits can be set by the potential measurements of even higher energy photon events with HAWC \cite{HAWC-ICRC} or CTA Telescopes. The latter, will have the potential to detect photons with energies up to 300 TeV, which could set constrains an order of magnitude higher.   


\section{Conclusions}

We have applied the LIV generic correction to photons to find the photon emission and decay rates. Due its role as an energy loss mechanism and the effects showed in this paper, these processes can lead to interesting searches of LIV-signatures in cosmic and gamma rays.

The generic rates reproduce characteristics and results based on the SME in the photon sector. Therefore they can provide highlights for further phenomenological studies
We have shown that LIV can motivate a trend to heavier cosmic ray primaries in the UHECR spectrum, inside an energetic window compatible with the UHECR regime, since the vacuum Cherenkov radiation threshold changes with the primary cosmic ray mass and charge. Up this energy region, heavier cosmic ray primaries will be attenuated faster that lighter due this mechanism.

Finally, when the achieved approximation is applied to photon decay process,  the rate is very efficient and highly constrain the free photon propagation over a given threshold, thus, photon decay can be used  as a direct and very simple way to bound the LIV energy scale by the direct observation of the very high energy photons.

{\small{

\bibliography{bibliography}

}}
\end{document}